# Demystifying the oracle: A "20 Questions" game to promote AI ethics and literacy

Eugenio TUFINO

*Dipartimento di Scienze Fisiche, Informatiche e Matematiche, Università di Modena e Reggio Emilia*
eugenio.tufino@unimore.it

**Abstract.** As Generative AI becomes a key component in physics education, a significant ethical challenge has emerged: the tendency of students to anthropomorphize Large Language Models (LLMs), treating them as authoritative "oracles" that retrieve fixed facts from an internal database. However, LLMs operate fundamentally as probabilistic engines. This paper describes the design and implementation of a didactic activity, a reduced version of the "20 Questions" game, aimed at making this stochastic nature directly observable. Unlike a human player who fixes a target object at the start of the game, students discover that the model generates answers based solely on local coherence with the interaction history. By utilizing functionalities such as re-sampling and history rewinding, students act as experimenters, observing how identical interaction histories can yield diverging narrative paths. We discuss how mapping these behaviors to familiar physics concepts provides the epistemic scaffolding necessary to promote informed skepticism, framing the verification of AI outputs not merely as a compliance rule, but as a technical necessity derived from the system's probabilistic nature.

*t*

## Introduction

In recent years, the use of Large Language Models (LLMs) in high school and undergraduate courses has become pervasive. These systems are often presented to and perceived by students as authoritative agents retrieving facts from an internal database. This anthropomorphic projection poses a significant ethical risk [1], not oracles. They do not query a structured knowledge base; instead they generate responses by sampling from a probability distribution conditioned on the input prompt and the preceding interaction history. Making this stochastic nature visible and intelligible to students is a non-trivial educational challenge, as standard chat interfaces are typically designed to mask uncertainty and present a coherent, single stream of text.

Building on recent works that use interactive computational activities to reveal the internal structure of language models [2, 3], we propose a complementary, phenomenological approach accessible to all students.

Inspired by Murray Shanahan's framing LLMs as stochastic simulators rather than symbolic reasoners [4], we use a reduced version of the "20 Questions" game to expose the model's non-deterministic mechanics. The choice of this specific game is motivated by a critical epistemic distinction highlighted by Shanahan regarding games of incomplete information. In a traditional game played by a human, the answerer fixes a specific object (the "ground truth") in their mind at the beginning of the game. Conversely, an LLM playing this game without an external constraint does not "think" of an object. It simply simulates the role of a player. Consequently, the "secret object" is not fixed at initialization but remains in a state of underdetermination. This dynamic mirrors the "surprise version" of the 20 Questions game described by John Archibald Wheeler [5], where reality emerges constructively from the questions asked.

In this paper, we describe the implementation of this game as a web application on Hugging Face [6], a widely used online platform for developing and sharing machine-learning models and applications. By allowing students to "rewind" and "re-sample" the dialogue, the tool demonstrates how small fluctuations can lead to macroscopic bifurcations (e.g., the same history leading to different objects). This visualizes the intrinsic uncertainty of AI, providing an intuitive bridge to physics concepts such as sensitivity to initial conditions and measurement statistics.

**Design and implementation**

The web-based application implements a simplified version of the classic guessing game. The game follows a standard format familiar to students: the user asks questions to identify a secret entity, and the system answers with "Yes" or "No". However, to fit the activity within a standard lesson timeframe and to reduce complexity, we introduced specific constraints:

Limit: The game is limited to 10 questions instead of 20.

Domain: The target is restricted to physical objects (excluding abstract concepts; while living beings are not strictly forbidden, the prompt naturally biases the generation towards non-living matter).

Feedback: Each "Yes/No" answer is accompanied by a short justification to provide context.

Significantly, the prompt design is intentional and central to the didactic goal. The system instructions specify that the model is "thinking of an object," yet no specific object is fixed in the prompt or the system memory at initialization. The LLM is, therefore, free to generate answers that are locally coherent with the interaction history up to that point, without being globally constrained by a pre-determined ground truth. This design choice ensures that the "secret object" functions as an emergent property of the conversation trajectory rather than a hidden variable to be uncovered (see Supplementary Material for the full system prompt).

The user interface was designed to expose the probabilistic mechanics usually hidden behind standard chat interfaces. A key design choice is the separation of the activity into distinct phases. The experimental controls are explicitly hidden during the gameplay phase to preserve the "illusion" of determinism, and only revealed later.

Phase 1: Linear Interaction (Line A). Initially, the interface presents a standard chat view (Figure 1). The student interacts linearly, asking questions to narrow down the domain. The user can choose the Temperature (T) via a slider to adjust the variance of the responses. In this phase, the Resample function is disabled. The student builds a unique history until they successfully "Guess" a compatible object or exhaust the 10 questions.

Phase 2: Transition (Rewind Selection). Once Phase 1 concludes, the interface unlocks the "Laboratory" mode. A dropdown menu allows the user to select any previous question $Q_k$ from the history as a bifurcation point.

Phase 3: Stochastic Exploration (Line B). Upon activating the Restart from Question $Q_k$ command, the system resets the state to the antecedent history $H_{k-1}$. Formally, this is defined as the sequence of interaction pairs recorded prior to step k:

$$H_{k-1} = \{(Q_1, A_1), \ldots, (Q_{k-1}, A_{k-1})\} \qquad (1)$$

This effectively cuts the conversation tree, opening a new branch ("Line B"). Only in this phase do the advanced experimental controls become active (Figure 2):

- Resampling (Stochastic Rerun): A dedicated button appears. It allows the user to re-submit the *same* question $Q_k$ multiple times against the *same* antecedent history $H_{k-1}$. This allows students to directly observe the probability distribution $P(\text{answer}|H_{k-1}, Q_k)$ and visualize potential flips (Yes →No) without advancing the question counter.

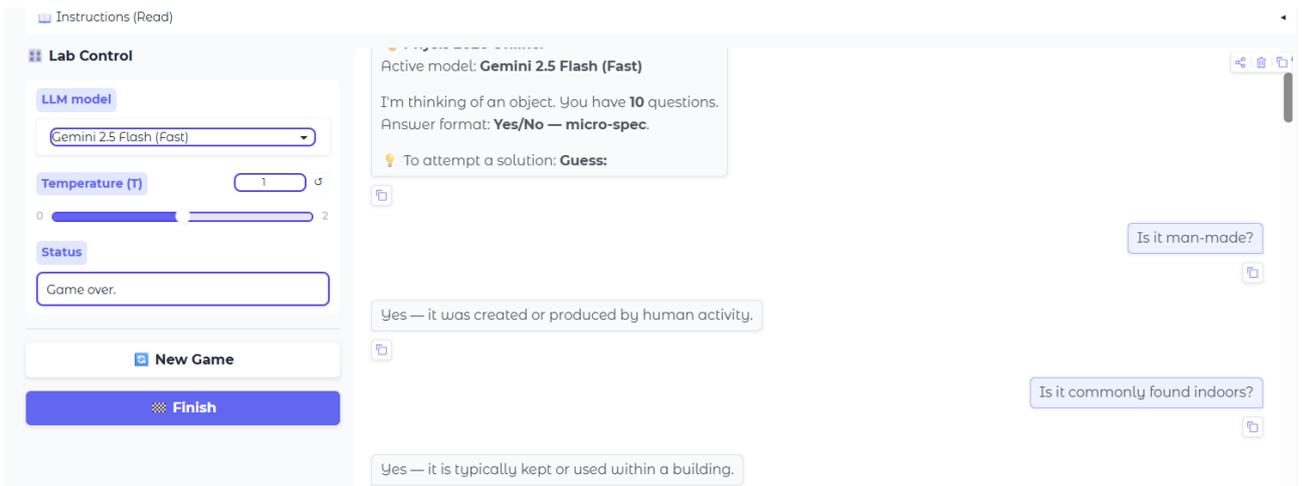

Figure 1. The interface during Phase 1. The student asks questions to build the history. The experimental controls (Restart/Resample) are hidden to insist on standard gameplay flow. User can choose the LLM model and the temperature parameter.

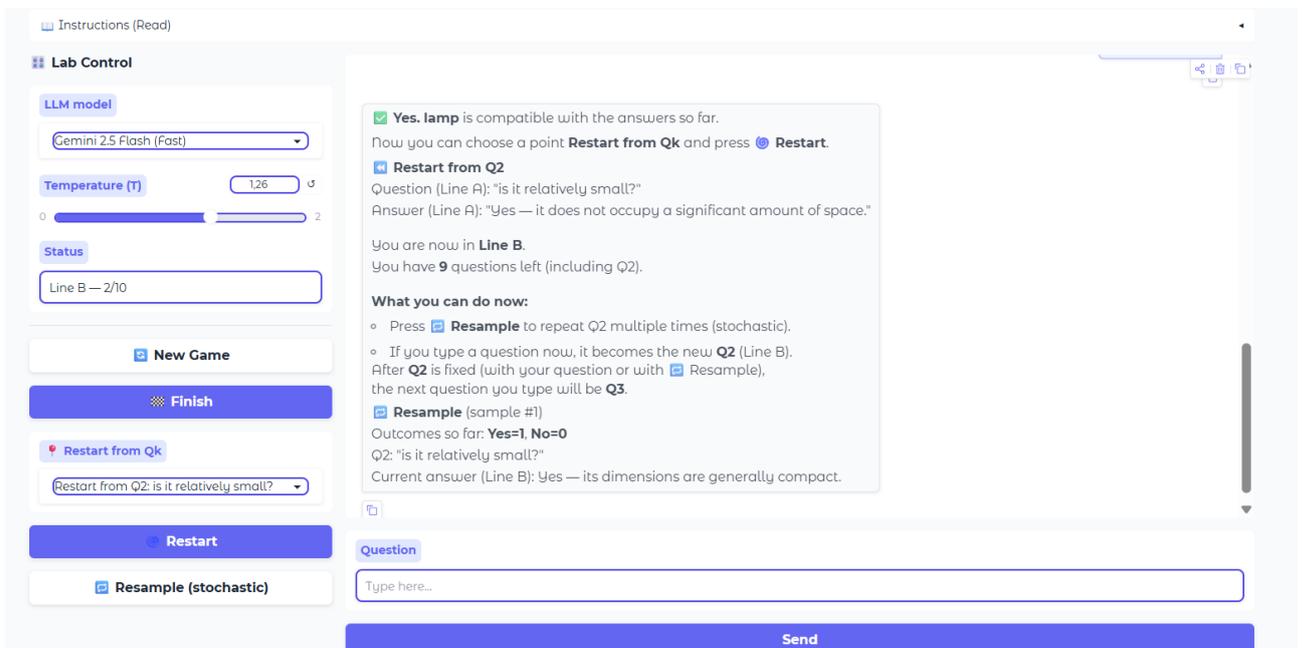

Figure 2. The interface during Phase 3. After restarting from Q2 , the Resample button becomes visible, allowing the user to repeatedly test the stability of the model's answer before committing to a new trajectory.

*Technical Implementation*

The application is implemented as a Python-based web app hosted on Hugging Face Spaces. A companion Google Colab Notebook is also provided, reproducing the core functionality [7]. From a technical perspective, the interaction is stateless at the LLM level: the model does not retain memory across API calls. The "illusion" of continuity in commercial ChatBot is created by the client application, which stores the dialogue history and re-sends the relevant context with each request. Details on access, source code availability, and deployment are reported in the Supplementary Material.

## Visualizing the stochastic nature of LLMs: A Case Study

To make the stochastic and non-committal behaviour of the LLM visible, we report a conversation sequence taken directly from the application interface (see Supplementary material for the whole sequence) with the choice of Google's Gemini 2.5 Flash model. This case study isolates a specific phenomenon: a stochastic flip under identical history via resampling, leading to a macroscopic bifurcation. In the initial linear playthrough (Line A), the user establishes a context of "man-made" objects $Q_1$. When asked $Q_2$ ("Is it relatively small?"), the model answers "Yes". This trajectory eventually leads to the compatible guess "Lamp". To test the stability of this outcome, the user utilizes the Restart function to return to $Q_2$ while preserving the antecedent history (see Figure 2). The user then employs the Re-sample function to repeat the same question multiple times against the fixed history ($H_1$).

As reported in Supplementary material, the resampling process yielded a dominant "Yes" trajectory (14 samples) and a rare "No" trajectory (2 samples). Justifications varied from "compact enough to be carried*"* (Yes) to "typically too large" (No). This observed variance confirms that the model does not hold a fixed internal truth, but samples from a probability distribution where the 'No' token retains a non-zero likelihood. Once the rare "No" outcome is obtained (Sample #16), the user selects it as the actual history for the new branch (Line B) and proceeds. The constraint set has now changed: the object is defined as "man-made" but not "relatively small".

The user asks subsequent questions: "Is it fixed in place?" ($\rightarrow$ No), "Is it used for transporting people?" ($\rightarrow$Yes), and "Does it have four wheels?" ($\rightarrow$ Yes). This new trajectory leads to the compatible guess "Car". The application displays an explicit bifurcation summary: Line A ends at "Lamp"; Line B ends at "Car".

This sequence isolates a single stochastic event (the flip on "relatively small") and demonstrates how it redirects the space of compatible continuations. To model this behaviour physically, we can describe the model's output not as a retrieved datum, but as a random variable a sampled from a conditional probability distribution [8]:

$$a \sim P(a|H, q, T) \qquad (2)$$

where a represents the generated answer (e.g., "Yes" or "No"), dependent on the interaction history H, the current question q, and the temperature T, which governs the level of randomness. This formalism clarifies that for a fixed history and question, the answer is not unique; students operationally experience that the "trajectory" is not pre-written but constructed dynamically through a stochastic process.

*Discussion and Educational Implications*

The phenomenology described in the previous section—specifically the observation of variance in sampling and the subsequent trajectory bifurcation—serves as scaffolding to discuss the epistemic nature of Generative AI. By engaging with the tool, students and teachers encounter behaviors that directly challenge the deterministic "oracle" view often associated with LLMs.

From an educational perspective, the web-based micro-laboratory maps effective analogies to concepts already familiar to physics students.

- Measurement noise and probability: The Re-sample function allows users to treat the model's output not as a fixed answer, but as a sample from a probability distribution. Repeatedly asking the same question (as in the sequence of 16 samples for $Q_2$) is analogous to collecting repeated measurements of a physical quantity affected by experimental uncertainty. Similarly, here users observe that for ambiguous questions (like "Is it small?"), the answers fluctuate between "Yes" and "No", revealing a broad distribution. Conversely, for strictly unambiguous properties (e.g., "Is it man-made?"), the result is more stable. This effectively visualizes the intrinsic uncertainty of the generative process.

- Sensitivity to Initial Conditions: The Restart mechanism highlights how the final system state (e.g., Lamp vs. Car) is highly sensitive to the interaction history. This offers a direct parallel to chaotic systems or diffusive processes, where a small fluctuation in the early stages (a single bit flip from "Yes" to "No") leads the system trajectory into a completely different region of the semantic space. The experiment demonstrates that the "object" is not a pre-existing hidden variable waiting to be discovered, but rather the result of a path-dependent evolution.

- Plausibility over truth: if the system can "invent" a car just to satisfy a probability distribution, it can equally fabricate a non-existent bibliography or an incorrect physical law. This realization provides students with a tangible reason to question the machine's authority, creating a necessary cognitive safeguard against automation bias in their coursework.

*Teacher and Student Observations*

During preliminary workshops with in-service teachers, a recurring question from teachers after observing the first bifurcations was: "How does the model decide which object to pick at the beginning?" This question marks a critical pedagogical entry point, revealing the persistence of the "Hidden Variable" misconception—the assumption that the AI, like a human player, holds a specific secret object in memory (at time t=0). The activity forces a shift in mental models. By observing that the system can legitimately "become" different objects (e.g., a Lamp or a Car) starting from the exact same interaction history, learners are induced to abandon the Retrieval Model (looking up a fact) in favor of a Generative Model. They recognize that the "decision" is merely a sampling event and that the "truth" of the object is emergent rather than intrinsic.

A key educational goal is to help students reconcile the variability observed in this game with the apparent stability of LLMs when solving standard textbook physics problems. The difference lies in the density of constraints. A recurring question from students during the activity was: if the model is inherently stochastic, why does it consistently converge on the correct solution for standard physics exercises? The key difference is how tightly constrained the task is:

- In the Game (underdetermined context): The prompt ("I am thinking of an object") imposes weak constraints. Consequently, the probability distribution over valid next tokens is broad, making stochastic fluctuations visible even at moderate temperatures.

- In Physics Problems (well- defined /highly constrained context): When a student asks an LLM to solve a specific physics problem, the rules of mathematical logic and physical laws act as powerful constraints. These shape the distribution such that the probability mass is concentrated on a narrow path of "correct" tokens [9]. This comparison clarifies that the underlying mechanism—next-token sampling—is identical in both cases. The model *appears* deterministic in physics tasks not because it switches to a "reasoning mode," but because the constraints render the "correct" path overwhelmingly more probable than the alternatives [10].

**Conclusions**

In this paper, we presented a reduced "20 Questions" game designed as a micro-laboratory to explore the stochastic nature of LLMs. This pedagogical activity allows students and teachers to verify also Murray Shanahan's theoretical framework: the LLM is not an agent accessing a hidden truth, but a stochastic simulator exploring a distribution of plausible continuations. The educational experience suggests that observing macroscopic bifurcations (such as the Lamp vs. Car case study) induces a necessary epistemic shift: users are compelled to abandon the naive deterministic view ("The model knows the answer") in favor of a probabilistic understanding ("The model samples a coherent narrative"). With a deeper understanding of LLM mechanics, this activity empowers educators to advocate for homework policies based on informed skepticism rather than abstract bans. For physics education, this approach is particularly valuable because it is based on students' existing scientific intuition. Concepts familiar to physics students—such as measurement statistics and sensitivity to initial conditions—become effective tools to understand better the behavior of Generative AI. It is important to note that frontier models can display behaviors that resemble reasoning or agency [11]. However, the variability highlighted by this activity follows directly from the probabilistic nature of generation and its sensitivity to the provided context. Future work will extend this investigation to standard physics problems, testing both classic textbook exercises and novel scenarios.

**Acknowledgments**

We gratefully thank Davide Riboli for interesting discussions on how to implement the 20 questions game.